# Large polaron evolution in anatase $TiO_2$ due to carrier and temperature dependence of electron-phonon coupling

**B.X. Yan[1,2], D.Y. Wan[1,2]*, X. Chi[2,3], C.J. Li[4], M.R. Motapothula[1], S. Hooda[1], P. Yang[3], Z. Huang[1], S.W. Zeng[1], A. Gadekar, S.J. Pennycook[4], A. Rusydi[1,2,3], Ariando[1,2], J. Martin[2], T. Venkatesan[1,2,4,5,6]***

*[1]NUSNNI-NanoCore, National University of Singapore, Singapore 117411, Singapore.*
*[2]Department of Physics, National University of Singapore, Singapore 117551, Singapore.*
*[3]Singapore Synchrotron Light Source, National University of Singapore, Singapore 117603, Singapore.*
*[4]Department of Material Science and Engineering, National University of Singapore, Singapore 117575, Singapore.*
*[5]NUS Graduate School for Integrative Sciences and Engineering, National University of Singapore, Singapore 117456, Singapore.*
*[6]Department of Electrical and Computer Engineering, National University of Singapore, Singapore 117583, Singapore.*
**Corresponding author. Email: Venky@nus.edu.sg; Wandy@nus.edu.sg*

**Abstract:**

Electronic and magneto transport data of anatase $TiO_2$ are analyzed and modelled considering various polaronic effects. The unexpected positive dependence of mobility on carrier concentration is attributed to the screening of electron-phonon interaction by excess charge carriers. The linear and quadratic components in magnetoresistance are separately associated with the transport and trapping modes of the carrier. From those results, we can depict the evolution of large polaron versus temperature and carrier concentration, and clarify the physics behind ambiguous conduction behavior of anatase $TiO_2$.

**Introduction:**

The anatase structural phase of titanium dioxide ($TiO_2$) is one of the most widely applied semiconductors. Besides its famous photosynthetic ability, novel applications rely on the less well-known electronic properties were proposed recently, such as memristors[1], spintronic devices[2], transistor[3] and transparent conducting oxides[4]. Stoichiometric anatase is an insulator with the band gap around 3.2 eV. The conductive samples, with intrinsic or extrinsic dopants, typically present with a carrier concentrations above $10^{17}$ $cm^{-3}$ and a resistivity among 1 to $10^{-4}$ $\Omega$·cm [5-6].

The conductive behavior of anatase depends on temperature and carrier concentration under multiple mechanisms. From the view of lattice structure, anatase $TiO_2$ is a disorder system[7]. Below ~60K, variable range hopping (VRH)[8] dominates and the dopants act as both trapping sites and disorder impurities, giving rise to the semiconductor behavior. When the electrons are thermally excited to conduction band, the dopants will only serve as scattering centers, resulting a metallic behavior. On the other hand, from the charge carrier perspective, electron in anatase $TiO_2$ would couple to phonon, trap itself and form the quasiparticle "large polaron"[9-11], an intermediate state between localized small polaron and free electron. Since the electron-phonon (e-ph) interaction is just moderate strong, the large polaron is easily perturbed and tuned to different states like dissociation[12-13], melting[14-15] and crystallization[16]. However, although extensive theoretical work has been performed, due to the lack of experimental methods, tentative characterization on the coupling range have only been launched recently with instruments like ARPES[10], RIXS[11] and STM[7],. Thus, few consistent description has yet been made on the polaron evolution, being a bottleneck for further research.

In this letter, we report on the polaron evolution in anatase $TiO_2$ over temperature and carrier concentration. Electronic and magnetic transport data from anatase $TiO_2$ thin films are used for exploring the polaron states. At the low and high temperature ranges, the resistivity shows semiconductor and metallic behavior, respectively, in agreement with previous studies. Unexpectedly, the mobility was found to increase with increasing carrier concentration. In the semiconductor phase, the higher hopping mobility results from an increase in hopping sites concentration. While, in the metallic phase, the increased mobility is attributed to a decrease in polaron effective mass due to a recently reported screening effect on e-ph coupling. We also find the magnetoresistance (MR) could be decomposed to a linear term and a quadratic term, separately characterizing the transport and trap behavior of carriers. Since different polaron states have different dependence on the temperature and carrier concentration, a phase diagram could be depicted through analyzation on the electronic and magneto transport data.

**Fabrication Methods:**

92nm (~100 monolayers) anatase $TiO_2$ thin films were grown on single crystal (100)-oriented $LaAlO_3$ substrates by pulsed laser deposition. All the samples were fabricated at 800°C under $3\times10^{-3}$ mTorr oxygen partial pressure. The laser energy and frequency was kept at 1.33J/cm$^2$ and 2 Hz. To reflect the nature properties of anatase $TiO_2$, we choose to introduce the electrons with the intrinsic dopant of oxygen vacancy by the vacuum thermal annealing. To avoid possible influence of annealing, we annealed all the samples at same temperature (900°C) for same time (3 hours). Doping level is only tuned with variation in oxygen partial pressure ( $9\times10^{-8}$ to $3\times10^{-3}$ mTorr). The diffusion length of oxygen ions $l = \sqrt{Dt}$ is 600nm, where D is the oxygen diffusion coefficient[17] and t is the annealing time. Since the diffusion length is much larger than the film thickness (~90nm),we can expect a homogenously distributed carrier concentration in the films.

**Results**

**Crystal Quality**

The X-ray diffraction (CuK$\alpha_1$ ray) pattern of the $TiO_2$ films is shown in Figure 1 (a). The sharp $TiO_2$ (004) peaks indicate the well-kept anatase structure. A tiny diffuse of the Bragg peak is shown in Fig1(b), indicating the lattice is slightly distorted after reduction. The distortion could be attributed to two possible mechanisms: (1) the oxygen vacancies can be viewed as flaws in the lattice frame, broadening the Bragg peak; (2) the e-ph interaction could distort the lattice, cause a diffusive Bragg peak[18]. The first mechanism is acceptable since all the FWHM ($\gamma$) of reduced samples are larger than the oxygen rich sample ($\gamma_0$), $\gamma/\gamma_0>1$. Nevertheless, it cannot explain the lattice relaxation at $\sim3\times10^{19}$cm$^{-3}$.which was reported as the criteria for a screening in e-ph coupling by excess carriers[10].Thus, the relaxation could be ascribed to the second mechanism: when the electron-lattice coupling is screened, the lattice will be relaxed from the distortion.

**Electronic transport measurement**

The transport properties of the anatase $TiO_2$ were measured by a Quantum Design Physical Properties Measurement System. The the resistivity of the samples range from $10^{-4}$ to $10^{-1}$ $\Omega\times$cm [Fig.2(a)], covering most reported data[6]. The Hall mobility and carrier concentration is presented in Fig.2(b). Unexpectedly, with higher carrier concentration, the mobility is higher. In the low temperature range (<60K), the carrier transport is believed in variable range hopping mode[8], corresponding to the semiconductor behavior, with[19]:

$$\rho = \rho_0 e^{\left(\frac{T_0}{T}\right)^{\frac{1}{4}}} \quad (1)$$

where $\rho$ is the resistivity, T is the temperature, $\rho_0$ and $T_0$ are two parameter fitted by experimental data. In the hopping model, the higher mobility is easily understood with a increase in trapping sites density. While in the metallic phase, the large polarons are excited to the conduction band, moving coherently and scattered by LO phonons, with the mobility $\mu_{LO}$[20]:

$$\mu_{LO} = \frac{\hbar}{2\alpha\hbar\omega_{LO}} \cdot \frac{e}{m_{LP}} \cdot \left(\frac{m_e^*}{m_{LP}}\right)^2 \cdot f(\alpha)\left(\mathrm{e}^{\frac{\hbar\omega_{LO}}{k_B T}} - 1\right) \quad (2)$$

where $e$ is electron charge, $\hbar$ is the reduced Plank's constant, $\omega_{LO}$ is the frequency of LO phonon involved in scattering, $\alpha$ is the electron phonon coupling constant, $m_e^*$ is the band effective mass of the carrier, $m_{LP} = m_e^*(1 + \alpha/6)$ is the effective mass of the large polaron, f(α) is a slowly varying function ranging from 1.0 to 1.4, and longitudinal optical frequency of anatase $TiO_2$ is 366$cm^{-1}$ [21]. The resistivity is fitted with the VRH model and the polaron model [Fig2(c)]. The calculated effective mass decrease with increasing carrier concentration, droping from 0.63$m_e$ at 7×$10^{18}$$cm^{-3}$ to 0.47$m_e$ at 2×$10^{20}$$cm^{-3}$, consistent with previous experimental result[10] for low $n_e$ (~0.7$m_e$ at 5×$10^{18}$$cm^{-3}$) and theoretical calculation in high $n_e$ ($m_e^*$ =0.43$m_e$ for bare effective mass)[22]. Thus, the positive dependence of mobility on carrier concentration could be explained with the screening effect of the large polaron. Namely, as the carrier concentration increases, the e-ph interaction is gradually screened[10]. Since the effective mass of a large polaron is proportional to its self-trapping potential[23] ,and carrier mobility is inverse proportional to the effective mass, the mobility will increase as the e-ph coupling is weaken.

**Magneto transport measurement**

The temperature dependent MR are shown in Fig3(a). With the decrease of temperature, the MR gradually change from positive to negative. In the low temperature range (<60K), MR can be fitted with a superposition of a negative linear term and a quadratic term. The negative component has been widely studied, resulting from the quantum interference effect[19]. This negative MR happens when the carrier transport incoherently in an disorder system[24]:

$$ln\left(\frac{\rho(T,H)}{\rho(T,0)}\right) = -A\nu\frac{1-\alpha}{\alpha}\left(\frac{e}{\hbar c}\cdot H\cdot n^{-\frac{2}{3}}\right)^{\frac{1}{2\nu}}\cdot\ln\left(\frac{\rho(T)}{\rho_0}\right)\ ,(\nu\sim 0.5) \quad (3)$$

where A is a dimensionless coefficient decided by experimental results, ν is a factor around 0.5 and n is the carrier concentration. In this equation, when the MR is small, namely, $\rho(T,H)/\rho(T,0)\sim 1$, the equation can be simplified to a linear format, corresponding to the negative linear component, with:

$$\frac{\Delta\rho}{\rho} = \beta_1 B \quad (4)$$

where B is the applied magnetic field and $\beta_1$ is the coefficient for the linear component. The quadratic term is attributed to the shrinkage of localized electronic wave function in the magnetic field. For a localized carrier, the magnetic field will be equivalent to an additional potential. This magnetic barrier will squeeze the electronic wave function, reduce the overlap between neighbors, cause a lower hopping rate, which result in a positive MR[25]:

$$\ln\left(\frac{\rho(H)}{\rho(0)}\right) = \frac{\tau a^* e^2 H^2}{N_i c^2 \hbar^2} \quad (5)$$

where τ is the scattering time, a* is the localization radius of the electronic wave function and $N_i$ is the doping level. When MR is small, the equation can be simplified to a quadratic format:

$$\frac{\Delta\rho}{\rho} = \beta_2 B^2 \quad (6)$$

where $\beta_2$ is the coefficient for the quadratic component.

In the metallic phase [Fig 4. (a)], as temperature drops, a transition from linear MR to quadratic MR is observed. Interestingly, we found it is also well fitted with a superposition of a linear part and quadratic part. However, MR behaviour in the high temperature is fewer focused than the low temperature, we need analyse the corresponding mechanism accordingly.

Two different models were proposed recently for the linear MR behaviour. The Quantum linear MR [26]is the one-landau-band limit of the SdH, with:

$$\frac{\Delta\rho}{\rho} = \frac{N_i\, H}{\pi n^2\, ec}, \left[n < \left(\frac{eH}{c\mathrm{h}}\right)^{\frac{3}{2}}\right] \quad (7)$$

For $n_e\sim 7\times 10^{18}cm^{-3}$, it requires B>20T, which is far beyond our measurement range. Besides, the linear MR here appears around room temperature, where quantum effects are usually eliminated by various scattering disturbance. Thus, this model is inconsistent with our experiment. The other model, classical linear MR[27], results from disorder scattering, with:

$$\frac{\Delta\rho}{\rho} \sim \frac{\pi\mu H}{2[g+0.35]} = \beta_1 B \quad (8)$$

where g is a numeric factor among 0 to 1. For the condition at 210K, where B=9T $\mu\approx 50cm^2/(V\times s)$ and MR~0.18%, the calculated g is acceptable, around 0.26. Moreover, a sign for the classical linear MR is its independence on carrier concentration, which is clearly shown in Fig 4. (b).

For a quadratic MR, the common explanation is the Lorentz motion[28] with:

$$\frac{\Delta\rho}{\rho} \sim (\mu B)^2 \tag{9}$$

For $\mu \approx 50 cm^2/(V \times s)$, B=9T, the MR is around 0.01%, which is much smaller than the experimental value ~0.2%. Such huge discrepancy indicates the Lorentz model is not applicable here. Noted that large polarons could protect the carrier from scattered by disorder[29], cooling will be beneficial for polaron formation, and the linearity of MR indeed decrease with dropping temperature, we assume that the MR transition from linear to quadratic is due to the formation of large polaron. When the polaron forms, the electron begins to self-trap. The magnetic field will add an additional magnetic potential to the system and perturb the electronic state[19], resulting in a positive quadratic MR, as:

$$m_{LP} = \frac{E_{LP}}{\omega_{LO}^2 a^{*2}}, V_H = \frac{\hbar^2 a^{*2}}{8m\lambda^4} \tag{10}$$

$$\rho(H) = \frac{m^*}{ne^2\tau} = \frac{E_{LP}(H)}{ne^2\tau \cdot \omega_{LO}^2 a^{*2}} = \frac{E_{LP}(0)+V_H}{ne^2\tau \cdot \omega_{LO}^2 a^{*2}} \tag{11}$$

thus:

$$\frac{\Delta\rho}{\rho} = \frac{V_H}{E_{LP}} = \left(\frac{e}{m}\right)^2 \frac{B^2}{8\omega_{LO}^2} = \beta_2 B^2 \tag{12}$$

where $\lambda = \sqrt{\hbar/(eB)}$ is called magnetic length. With B=9, the MR is around 0.26%, consisting with our experimental result.

**Disscussion**

During above discussion, we find: (1) a positive linear MR corresponds to the coherent motion of the carrier in a disorder system; while a negative MR indicates the incoherent hopping transport. (2) a small quadratic MR at high temperature results from the weak polaronic self-trapping; while a larger quadratic MR at low temperature represent a stronger defect trapping. Hence, the linear and quadratic terms can be interpreted as an indicator for the transport and trap condition of the carrier. Since different polaron state corresponds to different transport and trap state, MR can be used to infer the polaron evolution.

At high temperature, the thermal vibration would break the e-ph coupling, resulting in a dissociation state[12-13].The electrons would be scattered by disorder and MR would be linear and independent of carrier concentration. In this state, when temperature decreases, the e-ph scattering are weaken, the mobility will increase as shown in Fig 2 (B), leading to an increasing $\beta_1$ .Besides, with attenuated thermal vibration, the polaronic effect would be strengthened, resulting in an increasing $\beta_2$. These dissociation behaviours can be observed at the range from 210K to 300K in Fig.5.(a).

As the temperature keeps dropping ,the system would finally transient to a typical large polaron system[23].Quadratic MR would dominate in this phase. And the linear coefficient $\beta_1$ would decrease since the formation of large polaron can protect the electron from scattered by disorder[29]. Typically, at the low carrier concentration limit, the force among polaron is repulsive[16].The polaron tend to be localized and hamper the increase of the mobility. However, if carrier concentration so high that polarons start to overlap, the system would become polaron liquid[14-15]. In the liquid phase, the e-ph interaction is attenuated, the carriers are more free and prone to various scattering mechanism. Hence, contrary to typical large polaron system, the liquid phase mobility would increase with dropping temperature. In Fig 5 (a), the 90K to 210K data consistent with above description.

When the temperature is low enough, the carriers start to drop from the CB to the trapping sites[8], forming a localized polaron state[7]. The carrier transport type would transient from coherent transport to the incoherent hopping transport. The $\beta_1$ would change from positive to negative. Since the type of trapping change from the weak extensive spread self-trapping to a stronger localized defect trapping, a transient in $\beta_2$ should also be observed. Corresponding behaviour can be found at the temperature below 90K [Fig.5 (b)]. Besides, in beta_2 around 60K, the transition between self-trapping and defect trapping is blurred above $\sim 4\times10^{19}$cm$^{-3}$, where the lattice starts to be relaxed [Fig1.(b)], polarons start to overlap and screening effect emerges[10]. Thus, we would choose this carrier concentration as the boundary for the liquid phase in the polaron evolution diagram. [Fig 5.(c)]

**Conclusion**

In summary, anatase $TiO_2$ samples was fabricated, covering mostly reported resistivity range from $10^{-1}$ to $10^{-4}$ $\Omega\times$cm. A phase diagram for the polaron evolution was plotted through detailed analysis on the electro and magneto transport data. Our results filled the gap of study on large polaron evolution in antase $TiO_2$ and built a bridge to understanding various reported anatase $TiO_2$ related transport experiments. We proved that anatase can be smoothly tuned from a weakly correlated metal to polaronic semiconductor through controlling the polaron state. This can help to engineer the $TiO_2$ based devices in industrial application.

**Reference:**

1. Strukov, D. B.; Snider, G. S.; Stewart, D. R.; Williams, R. S., The missing memristor found. *Nature* **2009,** *459* (7250), 1154-1154.
2. Toyosaki, H.; Fukumura, T.; Yamada, Y.; Nakajima, K.; Chikyow, T.; Hasegawa, T.; Koinuma, H.; Kawasaki, M., Anomalous Hall effect governed by electron doping in a room-temperature transparent ferromagnetic semiconductor. *Nat Mater* **2004,** *3* (4), 221-224.
3. Wöbkenberg, P. H.; Ishwara, T.; Nelson, J.; Bradley, D. D. C.; Haque, S. A.; Anthopoulos, T. D., TiO2 thin-film transistors fabricated by spray pyrolysis. *Applied Physics Letters* **2010,** *96* (8), 082116.
4. Furubayashi, Y.; Hitosugi, T.; Yamamoto, Y.; Inaba, K.; Kinoda, G.; Hirose, Y.; Shimada, T.; Hasegawa, T., A transparent metal: Nb-doped anatase TiO2. *Applied Physics Letters* **2005,** *86* (25), 252101.
5. Forro, L.; Chauvet, O.; Emin, D.; Zuppiroli, L.; Berger, H.; Lévy, F., High mobility n-type charge carriers in large single crystals of anatase (TiO2). *Journal of Applied Physics* **1994,** *75* (1), 633-635.
6. Atsushi, K.; Shigeru, K.; Takahito, T.; Kazuyoshi, Y.; Yoji, K.; Hiroshi, K., Highly Reduced Anatase TiO 2-δ Thin Films Obtained via Low-Temperature Reduction. *Applied Physics Express* **2011,** *4* (3), 035801.
7. Setvin, M.; Franchini, C.; Hao, X.; Schmid, M.; Janotti, A.; Kaltak, M.; Van de Walle, C. G.; Kresse, G.; Diebold, U., Direct View at Excess Electrons in ${\mathrm{TiO}}_{2}$ Rutile and Anatase. *Physical Review Letters* **2014,** *113* (8), 086402.
8. Zhao, Y. L.; Lv, W. M.; Liu, Z. Q.; Zeng, S. W.; Motapothula, M.; Dhar, S.; Ariando; Wang, Q.; Venkatesan, T., Variable range hopping in TiO2 insulating layers for oxide electronic devices. *AIP Advances* **2012,** *2* (1), 012129.
9. Jaćimović, J.; Vaju, C.; Magrez, A.; Berger, H.; Forró, L.; Gaál, R.; Cerovski, V.; Žikić, R., Pressure dependence of the large-polaron transport in anatase TiO 2 single crystals. *EPL (Europhysics Letters)* **2012,** *99* (5), 57005.
10. Moser, S.; Moreschini, L.; Jaćimović, J.; Barišić, O. S.; Berger, H.; Magrez, A.; Chang, Y. J.; Kim, K. S.; Bostwick, A.; Rotenberg, E.; Forró, L.; Grioni, M., Tunable Polaronic Conduction in Anatase ${\mathrm{TiO}}_{2}$. *Physical Review Letters* **2013,** *110* (19), 196403.
11. Moser, S.; Fatale, S.; Krüger, P.; Berger, H.; Bugnon, P.; Magrez, A.; Niwa, H.; Miyawaki, J.; Harada, Y.; Grioni, M., Electron-Phonon Coupling in the Bulk of Anatase ${\mathrm{TiO}}_{2}$ Measured by Resonant Inelastic X-Ray Spectroscopy. *Physical Review Letters* **2015,** *115* (9), 096404.
12. Holstein, T., Studies of polaron motion. *Annals of Physics* **1959,** *8* (3), 343-389.
13. Liu, W.; Li, Y.; Qu, Z.; Gao, K.; Yin, S.; Liu, D.-S., Effect of Temperature on Polaron Stability in a One-Dimensional Organic Lattice. *Chinese Physics Letters* **2009,** *26* (3), 037101.
14. Fratini, S.; Quémerais, P., Melting of a Wigner crystal in an ionic dielectric. *The European Physical Journal B - Condensed Matter and Complex Systems* **2000,** *14* (1), 99-113.
15. Alexandrov, A. S.; Devreese, J. T., *Advances in polaron physics*. Springer: 2010.
16. Rastelli, G.; Ciuchi, S., Wigner crystallization in a polarizable medium. *Physical Review B* **2005,** *71* (18), 184303.
17. Astier, M.; Vergnon, P., Determination of the diffusion coefficients from sintering data of ultrafine oxide particles. *Journal of Solid State Chemistry* **1976,** *19* (1), 67-73.

18. Vasiliu-Doloc, L.; Rosenkranz, S.; Osborn, R.; Sinha, S. K.; Lynn, J. W.; Mesot, J.; Seeck, O. H.; Preosti, G.; Fedro, A. J.; Mitchell, J. F., Charge Melting and Polaron Collapse in ${\mathrm{La}}_{1.2}{\mathrm{Sr}}_{1.8}{\mathrm{Mn}}_{2}{O}_{7}$. *Physical Review Letters* **1999,** *83* (21), 4393-4396.
19. Shklovskii, B. I.; Efros, A. L., *Electronic properties of doped semiconductors*. Springer Science & Business Media: 2013; Vol. 45.
20. Low, F. E.; Pines, D., Mobility of Slow Electrons in Polar Crystals. *Physical Review* **1955,** *98* (2), 414-418.
21. Gonzalez, R. J.; Zallen, R.; Berger, H., Infrared reflectivity and lattice fundamentals in anatase ${\mathrm{TiO}}_{2}$s. *Physical Review B* **1997,** *55* (11), 7014-7017.
22. Taro, H.; Hideyuki, K.; Koichi, Y.; Hiroyuki, N.; Yutaka, F.; Shoichiro, N.; Naoomi, Y.; Akira, C.; Hiroshi, K.; Masaharu, O.; Yasushi, H.; Toshihiro, S.; Tetsuya, H., Electronic Band Structure of Transparent Conductor: Nb-Doped Anatase TiO 2. *Applied Physics Express* **2008,** *1* (11), 111203.
23. Emin, D., *Polarons*. Cambridge University Press, New York: 2013.
24. Sladek, R. J., Magnetically induced impurity banding in n-InSb. *Journal of Physics and Chemistry of Solids* **1958,** *5* (3), 157-170.
25. Pollak, M.; Shklovskii, B., *Hopping transport in solids*. Elsevier: 1991; Vol. 28.
26. Abrikosov, A. A., Quantum magnetoresistance. *Physical Review B* **1998,** *58* (5), 2788-2794.
27. Parish, M. M.; Littlewood, P. B., Non-saturating magnetoresistance in heavily disordered semiconductors. *Nature* **2003,** *426* (6963), 162-165.
28. Abrikosov, A. A.; Beknazarov, A., *Fundamentals of the Theory of Metals*. North-Holland Amsterdam: 1988; Vol. 1.
29. Zhu, H.; Miyata, K.; Fu, Y.; Wang, J.; Joshi, P. P.; Niesner, D.; Williams, K. W.; Jin, S.; Zhu, X.-Y., Screening in crystalline liquids protects energetic carriers in hybrid perovskites. *Science* **2016,** *353* (6306), 1409-1413.

**Figure Caption**

Figure 1. (a) X-ray diffraction of as-grown and reduced TiO2 films on LAO substrate. (b) Normalized intensity of X-ray scattering around the (004) Bragg reflection. The FWHM ratio of the reduced to the oxygen-rich sample is drawn in the inset of (b).

Figure 2.(a) Temperature dependence of resistivity in anatase TiO2 thin film. (b) Carrier concentration and mobility measured by Hall effect. In 3-300K range, a higher carrier concentration always corresponds to a higher mobility. (c) Measured (symbols) and modelled (solid line) resistivity in the $7 \times 10^{18}\mathrm{cm}^{-3}$ sample. Variable range hopping and polaron scattering are included. (d) Calculatxd effective mass of the large polaron.

Figure 3. (a) Magnetoresistance (MR) in the $7 \times 10^{18}\mathrm{cm}^{-3}$ sample. (b)-(d) MR of samples with different carrier concentration at 60K, 10K and 3K (e) Experimental and fitting curves of MR for the $7 \times 10^{18}\mathrm{cm}^{-3}$sample at 60K,10K and 3K. The MR can be well decomposed to a negative linear component and a quadratic component. This applies to all the samples.

Figure 4. (a) Fine structure of MR in the $7 \times 10^{18}\mathrm{cm}^{-3}$sample in high temperature (>120K). (b)-(d) MR of samples with different carrier concentration at 300K, 210K and 120K (e) Experimental and fitting curves of MR for the $7 \times 10^{18}\mathrm{cm}^{-3}$sample at 300K ,210K and 120K. MR in high temperature can also be well decomposed to a linear component and a quadratic component.

Figure 5. (a)-(b) The temperature dependent of parameter $\beta_1$ and $\beta_2$ of the Anatase samples. (c) Phase diagram for large polaron in anatase TiO2, where we have sketched the proposed polaron phases and how it might evolve according to temperature and carrier concentration. The boundary 1 circle points are the maximum values of $\beta_1$, separating the dissociation phase from the polaronic phases. The boundary 2 square points are zero points of $\beta_1$, separating the localized state from the mobile states.

Fig.1

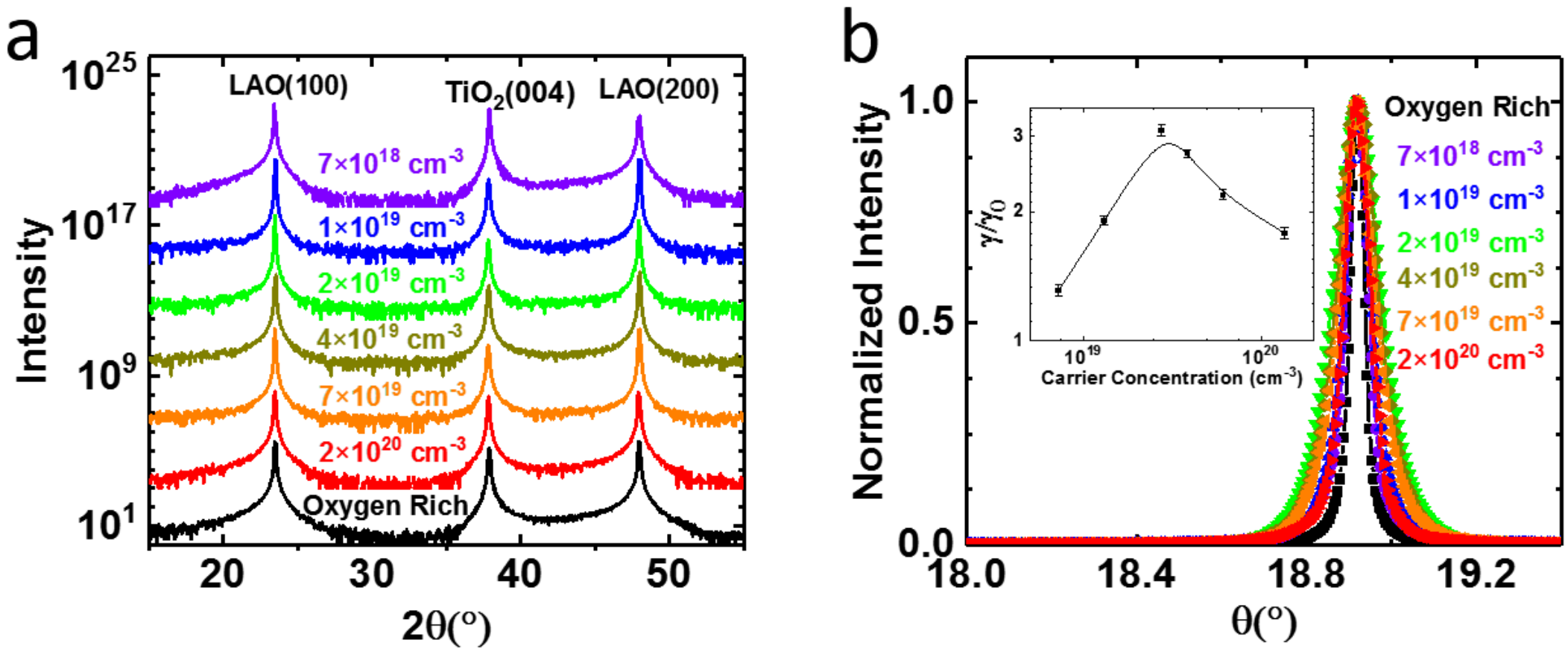

**Fig.2**

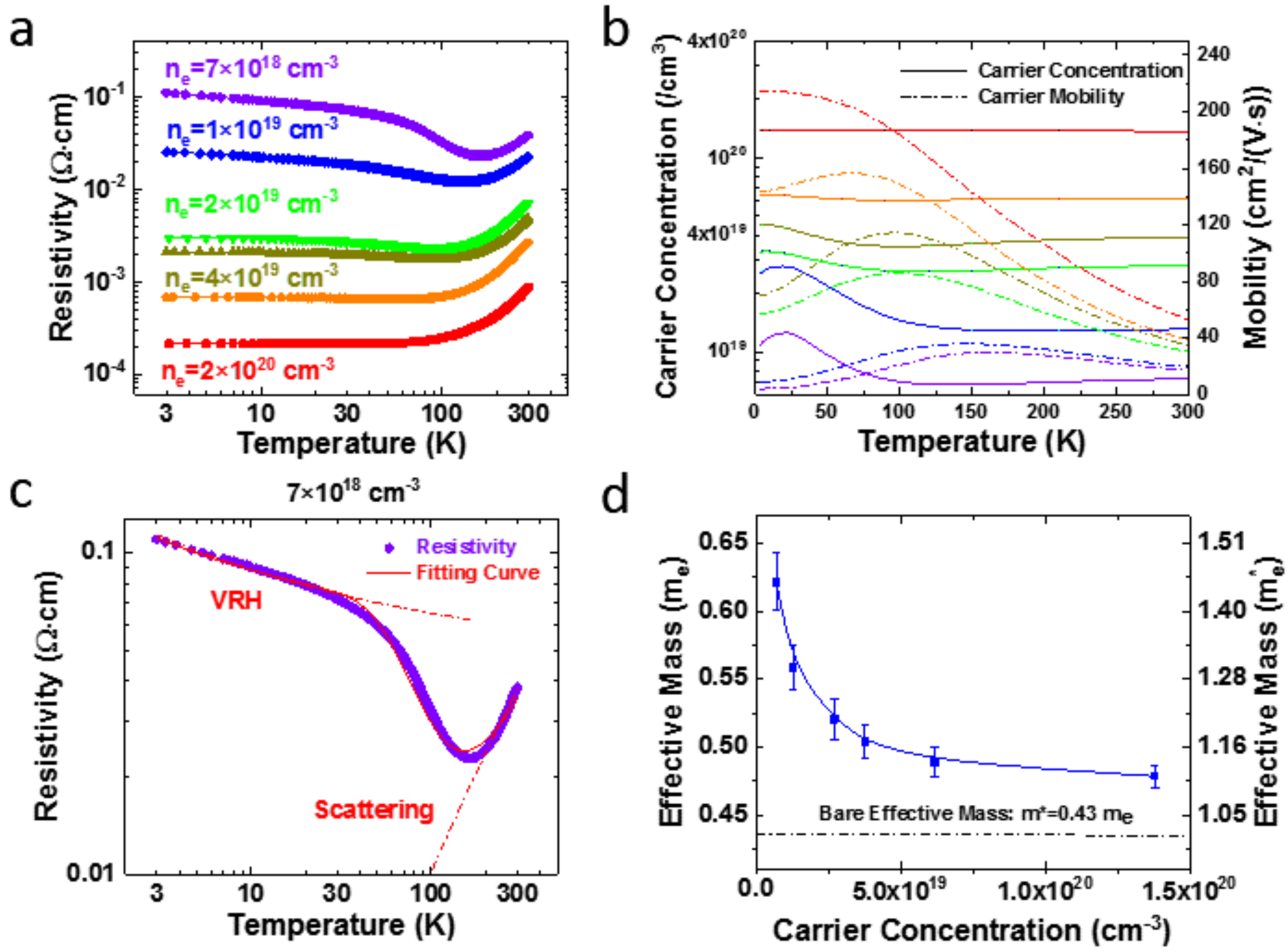

**Fig.3**

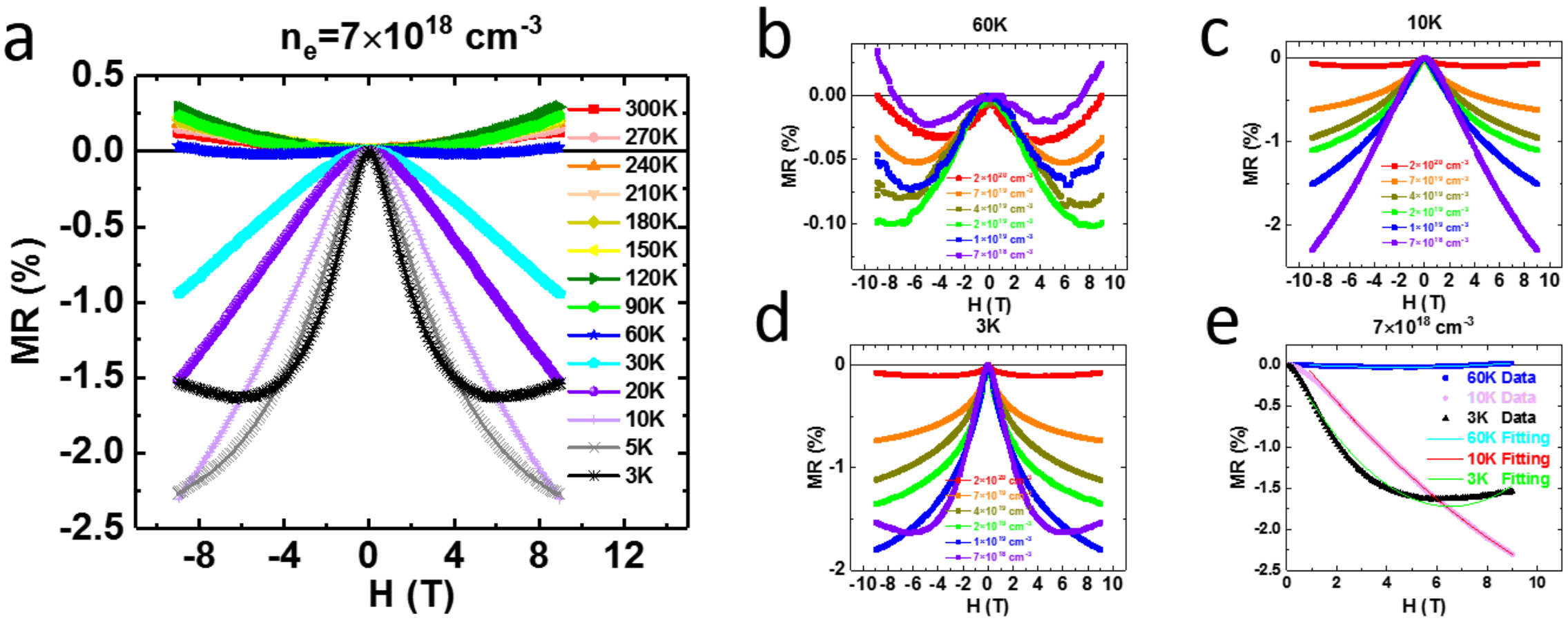

**Fig.4**

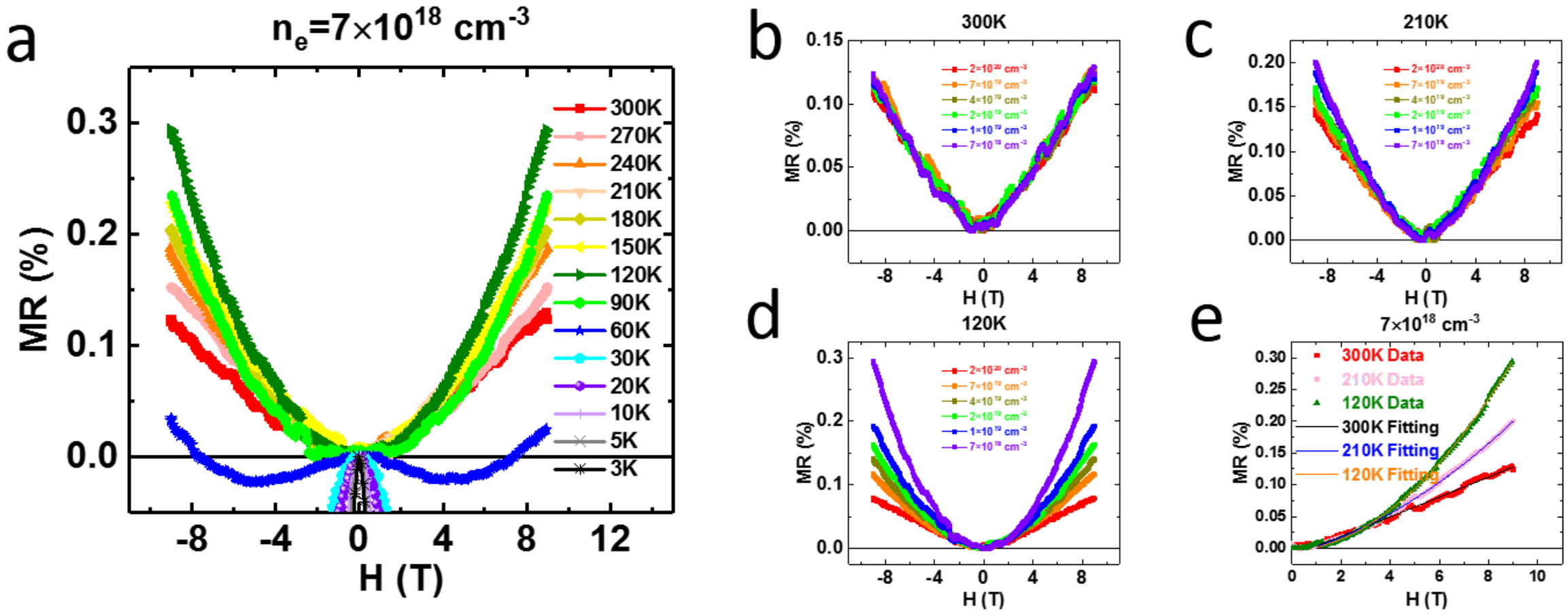

**Fig.5**

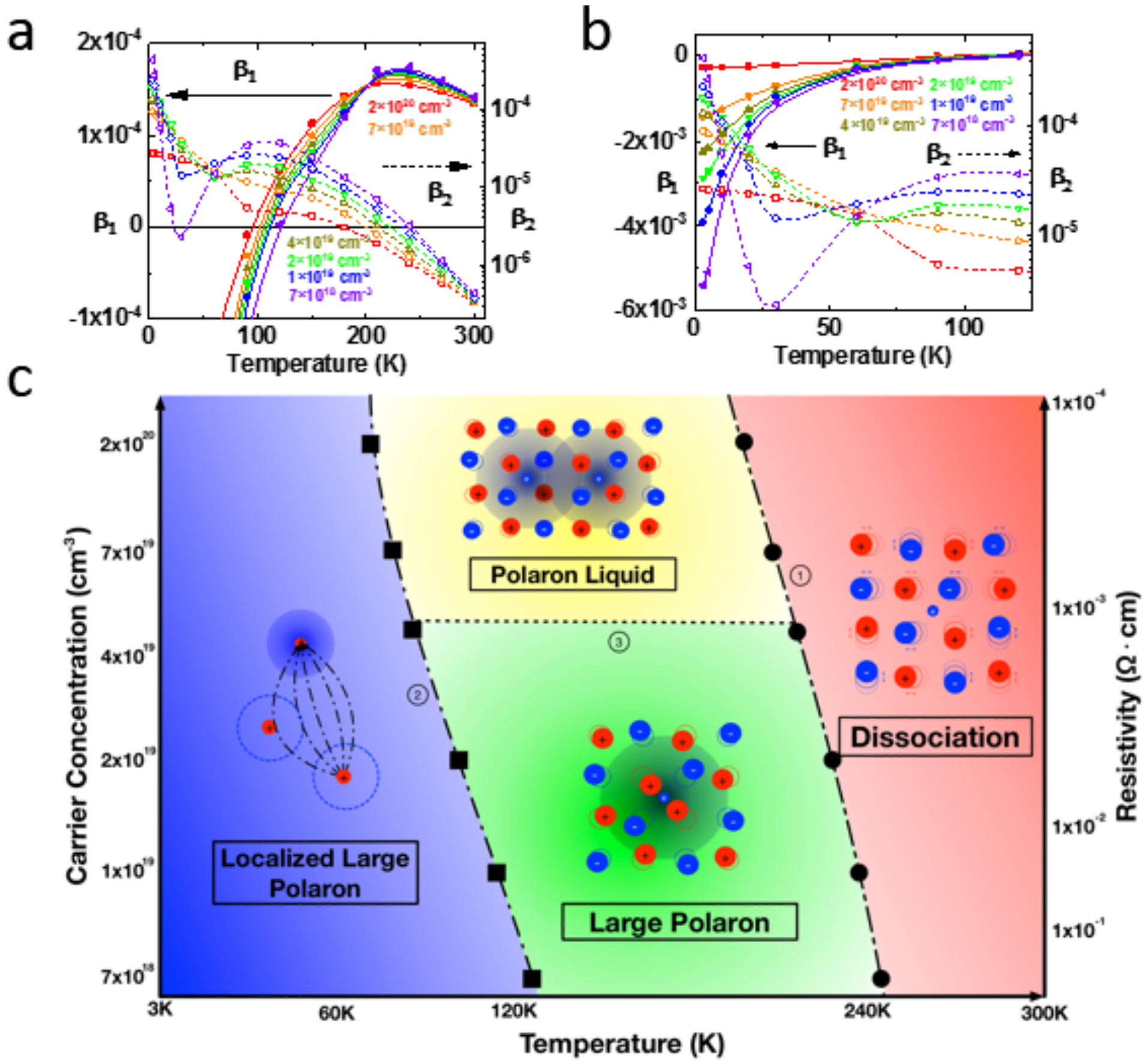